\RequirePackage{lineno}

\documentclass[twocolumn,english,aps,prc,showpacs,superscriptaddress,floatfix]{revtex4}
\usepackage[utf8]{inputenc} 
\usepackage{color}
\usepackage{multirow}
\usepackage{amsmath}
\usepackage{graphicx}
\usepackage{babel}
\usepackage[normalem]{ulem}

\newcommand {\snn}	{\sqrt{s_{_{\rm NN}}}}

\newcommand {\dvvach}	{\Delta v_{2}(A_{\rm ch})}
\newcommand {\dvvvach}	{\Delta v_{3}(A_{\rm ch})}
\newcommand {\dvnach}	{\Delta v_{n}(A_{\rm ch})}

\newcommand {\ach} {A_{\rm ch}}

\begin{document}
\title{Complications in the interpretation of the charge asymmetry dependent $\pi$ flow for the chiral magnetic wave}
\author{Hao-jie Xu}
\email{haojiexu@zjhu.edu.cn}
\affiliation{School of Science, Huzhou University, Huzhou, Zhejiang 313000, China}
\affiliation{Department of Physics and Astronomy, Purdue University, West Lafayette, Indiana 47907, USA}
\author{Jie Zhao}
\affiliation{Department of Physics and Astronomy, Purdue University, West Lafayette, Indiana 47907, USA}
\author{Yicheng Feng}
\affiliation{Department of Physics and Astronomy, Purdue University, West Lafayette, Indiana 47907, USA}
\author{Fuqiang Wang}
\email{fqwang@zjhu.edu.cn}
\affiliation{School of Science, Huzhou University, Huzhou, Zhejiang 313000, China}
\affiliation{Department of Physics and Astronomy, Purdue University, West Lafayette, Indiana 47907, USA}

\date{\today}

\begin{abstract}
The charge asymmetry ($A_{\rm ch}$) dependence of the $\pi^{-}$ and $\pi^{+}$ elliptic flow difference, $\Delta v_{2}(A_{\rm ch})$, has been regarded as a
	sensitive observable for the possible chiral magnetic wave (CMW) in relativistic heavy ion collisions. 
In this work, we first demonstrate that, due to non-flow backgrounds, the flow measurements by the Q-cumulant method using all charged particles as 
	reference introduce a trivial linear term to $\Delta v_{2}(A_{\rm ch})$. The trivial slope can be negative in the triangle flow difference $\Delta v_{3}(A_{\rm ch})$ 
	if the non-flow is dominated by back-to-back pairs.
	After eliminating the trivial term, 
	we find that the non-flow between like-sign pairs gives rise to an additional positive slope to $\Delta v_{2}(A_{\rm ch})$ 
	because of the larger dilution effect to $\pi^{+}$ ($\pi^{-}$) at positive (negative) $A_{\rm ch}$. 
	We further find that the competition between different $\pi$ sources can introduce another  non-trivial linear-$A_{\rm ch}$ term due to their different multiplicity 
	fluctuations and anisotropic flows.
	We then study the effect of neutral cluster (resonance) decays as a mechanism for local charge conservation on the slope parameter of  
	$\Delta v_{2}(A_{\rm ch})$. We find that the slope parameter is sensitive to the kinematics of those neutral clusters. 
	Light resonances give positive slopes while heavy resonances give negative slopes. Local charge conservation from continuum cluster mass distribution can give a positive slope parameter comparable to experimental data.
	Our studies indicate that many non-CMW physics mechanisms can give rise to a $A_{\rm ch}$-dependent $\Delta v_{2}(A_{\rm ch})$ and the interpretation of $\Delta v_{2}(A_{\rm ch})$ in terms of the CMW is delicate. 
\end{abstract}


\maketitle

\section{Introduction}

One of the most important phenomena observed in high-energy heavy ion collisions at the Relativistic Heavy Ion Collider (RHIC) 
and the Large Hadron Collider (LHC) is the strong collective flow of final state charged hadrons~\cite{Adams:2005dq,Adcox:2004mh,Arsene:2004fa,Back:2004je,Muller:2012zq}. This is developed, presumably, by fast expansion of the quark gluon plasma 
(QGP) created in those collisions~\cite{Ollitrault:1992bk,Kolb:2000fha}. 
The flow harmonics, such as elliptic flow ($v_{2}$) and triangular flow ($v_{3}$), are widely used to study and extract the transport properties
of the QGP~\cite{Molnar:2001ux,Petersen:2010cw,Gale:2012rq,Xu:2014ada,Song:2017wtw}.
One complication in those studies is the non-flow contaminations in the $v_{n}$ measurements~\cite{Borghini:2006yk,Ollitrault:2009ie,Eyyubova:2009hh,Agakishiev:2011eq,Xu:2012ue,Jia:2013tja,Abdelwahab:2014sge}.
These non-flow correlations can be caused by resonance decays, jet emission, etc. 
At low transverse momentum ($p_{T}$), however, non-flow contaminations in flow measurements are generally small.

Recently, pion elliptic flows have been  proposed as a tool to search for a phenomenon called the chiral magnetic wave (CMW)~\cite{Kharzeev:2010gd,Burnier:2012ae}, which is related to 
the chiral anomaly and strong magnetic field in heavy ion collisions~\cite{Kharzeev:2007tn,Kharzeev:2007jp,Fukushima:2008xe,Son:2009tf,Kharzeev:2010gd,Zhitnitsky:2010zx,Bzdak:2011yy,Burnier:2012ae,Deng:2012pc,Tuchin:2014iua,Copinger:2018ftr,Han:2019fce,Siddique:2019gqh}. 
The CMW is a gapless collective excitation of the QGP stemming
from the interplay of the chiral magnetic and chiral separation effects~\cite{Kharzeev:2007tn,Kharzeev:2007jp,Fukushima:2008xe,Kharzeev:2010gd,Zhitnitsky:2010zx,Burnier:2012ae}.
The CMW could introduce an electric quadrupole moment, 
giving opposite contributions to the $v_{2}$'s of positive and negative pions~\cite{Burnier:2012ae}. 
These contributions would depend on
\begin{equation}
	\ach=\frac{N_{+}-N_{-}}{N_{+}+N_{-}},
\end{equation}
where $N^{+}$ ($N^{-}$) is the multiplicity of positive (negative) charged hadrons in a given event. 
Namely, the CMW would cause the pion flows to linearly dependent on $\ach$,
\begin{equation}
	v_{2}\{\pi^{\pm}\} =  v_{2}^{\rm base} \mp \frac{r(\pi^{\pm})}{2}\ach,
\end{equation}
with opposite sign slope parameters for $\pi^{+}$ and $\pi^{-}$. The CMW-sensitive slope parameter ($r$) can be extracted from the $\ach$
dependent pion flow difference
\begin{equation}
	\dvvach =  v_{2}\{\pi^{-}\} - v_{2}\{\pi^{+}\}.
\end{equation}
Experimental measurements by the STAR, ALICE and CMS collaborations qualitatively agree with the expectation from the CMW~\cite{Adamczyk:2015eqo,Adam:2015vje,Sirunyan:2017tax}. 
Firm conclusions, however, have not been reached because it is generally perceived that non-CMW
mechanisms can also generate $\ach$-dependent $\pi$ flows, e.g, the Local Charge Conservation (LCC)~\cite{Bzdak:2013yla} and the effect of isospin chemical potential~\cite{Hatta:2015hca}.
Some of them are of non-flow nature, such as the LCC, but the effect of LCC on $\dvvach$ is not coming from non-flow. 
The key reason for the LCC effect on $\dvvach$ is that the charge asymmetry
is more effectively affected by the decay products of neutral clusters with lower $p_{T}$, resulting in a smaller $\pi^+$($\pi^-$) $v_{2}$ at positive (negative) $\ach$~\cite{Bzdak:2013yla}. 

Non-flow contributions to $\dvvach$ have not been thoroughly discussed in literature. Non-flow contributions may be small in the $\pi$ anisotropic flow measurements themselves,
but could be important in the small difference between $\pi^{+}$ and $\pi^{-}$ flows. We will demonstrate in this paper that
the non-flow correlations can give both trivial and non-trivial contributions to the slope parameters of $\dvnach$ in Sec.~\ref{sec:trivial}. 
We will then discuss the effect of multiple pion sources with different flows and contributions to the charge asymmetry in Sec.~\ref{sec:multiple}.
We will revisit non-flow correlations in Sec.~\ref{sec:likesign} and show that non-flow correlations can give further, non-trivial contributions to the slope parameters. 
We will further show in Sec.~\ref{sec:kinematics} and Sec.~\ref{sec:lcc} that the effect of LCC may not be as simple as it appears. It can give rise to both positive and negative slopes, depending the details of the LCC physics.
 We will illustrate that continuum mass distributions of charge conserving clusters may play an important role in $\dvvach$.

\section{Trivial linear term in $v_{n}(\ach)$}
\label{sec:trivial}

The Q-cumulant method was widely used to calculate anisotropic flow observables~\cite{Bilandzic:2010jr,Bilandzic:2013kga,Xu:2016hmp,Zhao:2017yhj}. 
The method is based on the $Q_{n}$-vector defined as 
\begin{equation}
Q_{n}=\sum_{i=1}^{M}e^{in\varphi_{i}},
\end{equation}
where $M$ is the multiplicity, $\varphi_{i}$ is the azimuthal
angle of the emitted particle $i$, and $n$ is the harmonic order.
Consider the case where there is no overlap between the particle of interest (POI) and the reference particle (REF), as is the case in our study,
the two-particle cumulant can be calculated by
\begin{equation}
	d_{n}\{2\}=\left\langle{\frac{q_{n}Q_{n}^{*}}{mM}}\right\rangle,
\end{equation}
where $(m,q_{n})$ and $(M,Q_{n})$ are the (multiplicity, Q-vector)
of POI and REF, respectively.
Here and hereafter, $\langle \rangle$ denotes average over events.
To reduce short range non-flow correlations in experimental studies, the sub-event method is often used with a minimum pseudo-rapidity gap $\Delta\eta$ 
applied between the POI and the REF. 
Following this common practice, we apply a gap of $\Delta\eta=0.6$ in our phenomenological study here, although non-flow is known in our study and a $\Delta\eta$ gap may not be necessary to remove it.
With all charged hadrons as REF, the anisotropic
flow of $\pi^{\pm}$ can be written as 
\begin{equation}
	v_{n}^{\pi^{\pm}}\{2\}=\frac{d_{2}\{2;\pi^{\pm}\mbox{-}{\rm REF}\}}{\sqrt{c_{2}\{2\}}}.
\end{equation}
Here we use $d_{n}\{2;A\mbox{-}B\}$
to represent the two-particle cumulant $d_{n}\{2\}$ between
A (e.g. POI) and B (e.g. REF). 
$c_{n}\{2\}\equiv d_{n}\{2;{\rm REF\mbox{-}REF}\}$ is that of the REF pairs, and $\sqrt{c_{n}\{2\}}$ is the reference flow.

We now demonstrate that, if including all charged particles (denoted as $h$) as REF, the above method introduces a trivial term linear in $\ach$.
Using $M=N_{+}+N_{-}$ and $Q_{n}=Q_{n+}+Q_{n-}$, where $N_{\pm}$
and $Q_{n\pm}$ are the multiplicities and Q-vectors of positive/negative
particles, the two-particle cumulant $d_{2}\{2;\pi^{\pm}h\}$~\footnote{Here we have omitted the dash between $\pi^{\pm}$ and $h$ which causes no confusion} can be rewritten as
\begin{align}
	& d_{2}\{2;\pi^{\pm}h\}  =\left\langle\frac{q_{n}^{\pi^\pm}Q_{n}^{*}}{mM}\right\rangle  \nonumber \\
	& = \left\langle\frac{N_{+}}{M}\frac{q_{n}^{\pi^\pm}Q_{n+}^{*}}{mN_{+}}\right\rangle +\left\langle \frac{N_{-}}{M}\frac{q_{n}^{\pi^\pm}Q_{n-}^{*}}{mN_{-}}\right\rangle  \nonumber \\
	& =\frac{d_{n}\{2;\pi^{\pm}h^{+}\}+d_{n}\{2;\pi^{\pm}h^{-}\}}{2} \nonumber \\ 
	&	+\frac{d_{n}\{2;\pi^{\pm}h^{+}\}-d_{n}\{2;\pi^{\pm}h^{-}\}}{2}\ach, \label{eq:trivial}
\end{align}
where we have used the fact that the event average $\langle\rangle$ is achieved
in a given $\ach$ interval.
A trivial term appears, which is proportional to $\ach$ (the second term on r.h.s of Eq.~\ref{eq:trivial}). If the correlations are from flow only, then $d_{n}\{2;\pi^{\pm}h^{+}\}=d_{n}\{2;\pi^{\pm}h^{-}\}$,
and the trivial term vanishes. However, non-flow is present in experimental data and 
differs between like-sign and unlike-sign pairs, so the trivial term is finite. 
It is opposite in sign for $\pi^{+}$ and $\pi^{-}$. The trivial term arises when all charged particles are included in REF and the non-flow differs between like-sign and unlike-sign
pairs.
The slope of the trivial term, dubbed the trivial slope $r_{\rm triv}$, reads,
\begin{equation}
	r_{\rm triv}(\pi^{\pm}) =  \frac{d_{n}\{2;\pi^{\pm}h^{+}\}-d_{n}\{2;\pi^{\pm}h^{-}\}}{2\sqrt{c_{n}\{2\}}}.
	\label{eq:rT}
\end{equation}
The slope parameter without removing the trivial term is denoted as $r_{0}$ in this paper.

We use a toy Monte Carlo (MC) model to illustrate the trivial term. 
We generate $\pi^{+}$ and $\pi^{-}$ with Poisson multiplicity fluctuations in each event. 
The $p_{T}$ of $\pi$ mesons correspond to the measured data in the $30\mbox{-}40\%$ centrality of Au+Au collisions at $\snn=200$ GeV~\cite{Zhao:2017nfq,Adamczyk:2015lme};
The $\eta$ spectra is parameterized as~\cite{Alver:2010ck} 
\begin{equation}
	\frac{dN_{\rm ch}}{d\eta}\propto\frac{\sqrt{1-1/(\alpha\cosh(\eta))}}{1+\exp((|\eta|-\beta)/a)},
	\label{eq:eta}
\end{equation}
with $\alpha=2.74$, $\beta=3.92$ and $a=0.88$ for the centrality of interest.
The mean multiplicity of charged hadrons is $380$ in $|\eta|<1$ with $p_{T}>0.15$ GeV/c.
To introduce a non-flow correlation difference between like-sign and unlike-sign pairs, we
force, on average, $20\%$ of the multiplicity in a given event to come from $\pi^{+}\pi^{-}$ pairs with identical azimuthal angle for the two pions.
A constant elliptic flow $v_{2}=4\%$ is used to generate the azimuth
angle of those pairs as well as the rest $80\%$ $\pi^{+}$ and $\pi^{-}$. 

\begin{figure*}[tbp]
\begin{centering}
\includegraphics[scale=0.3399]{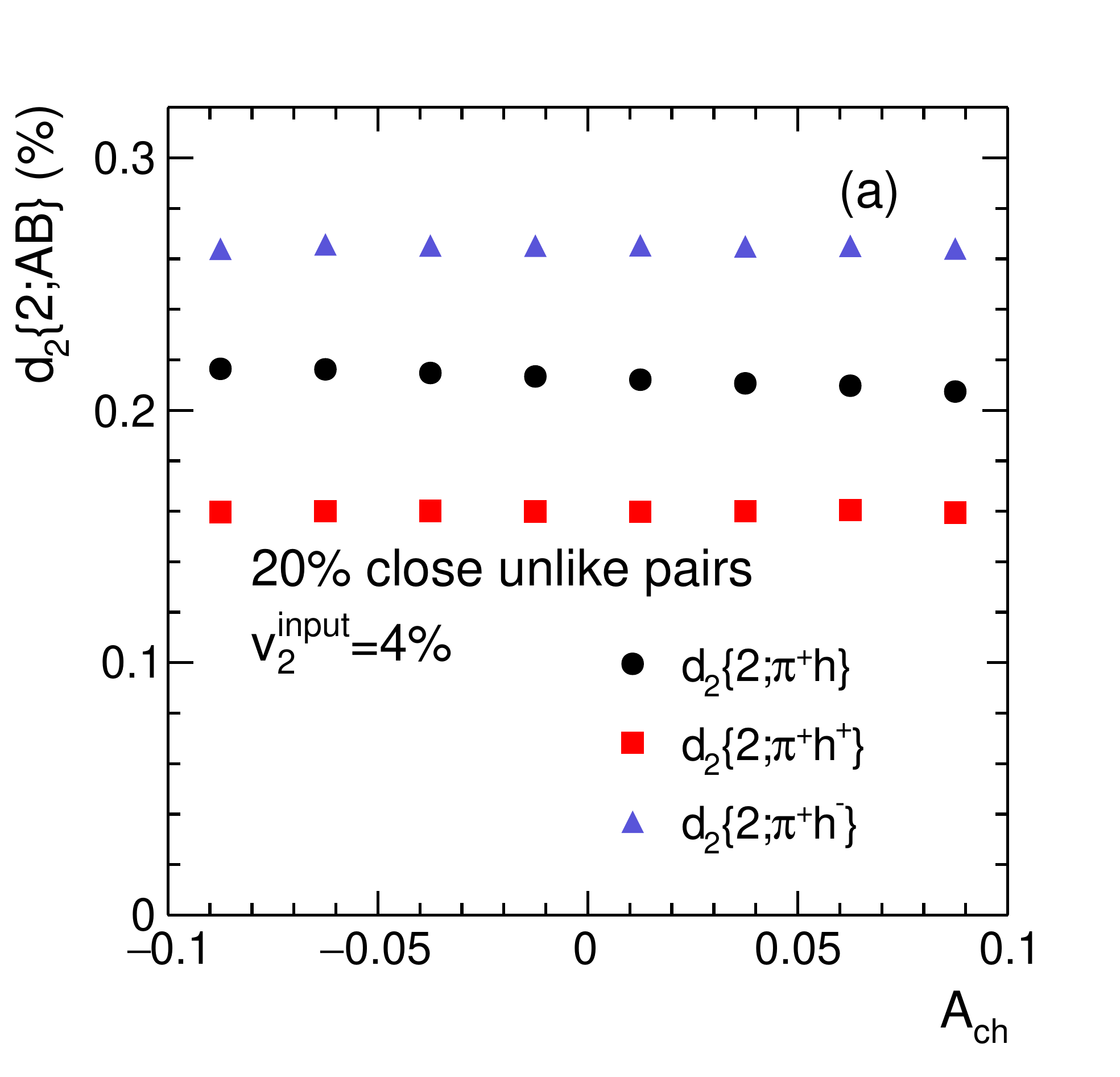}\includegraphics[scale=0.3399]{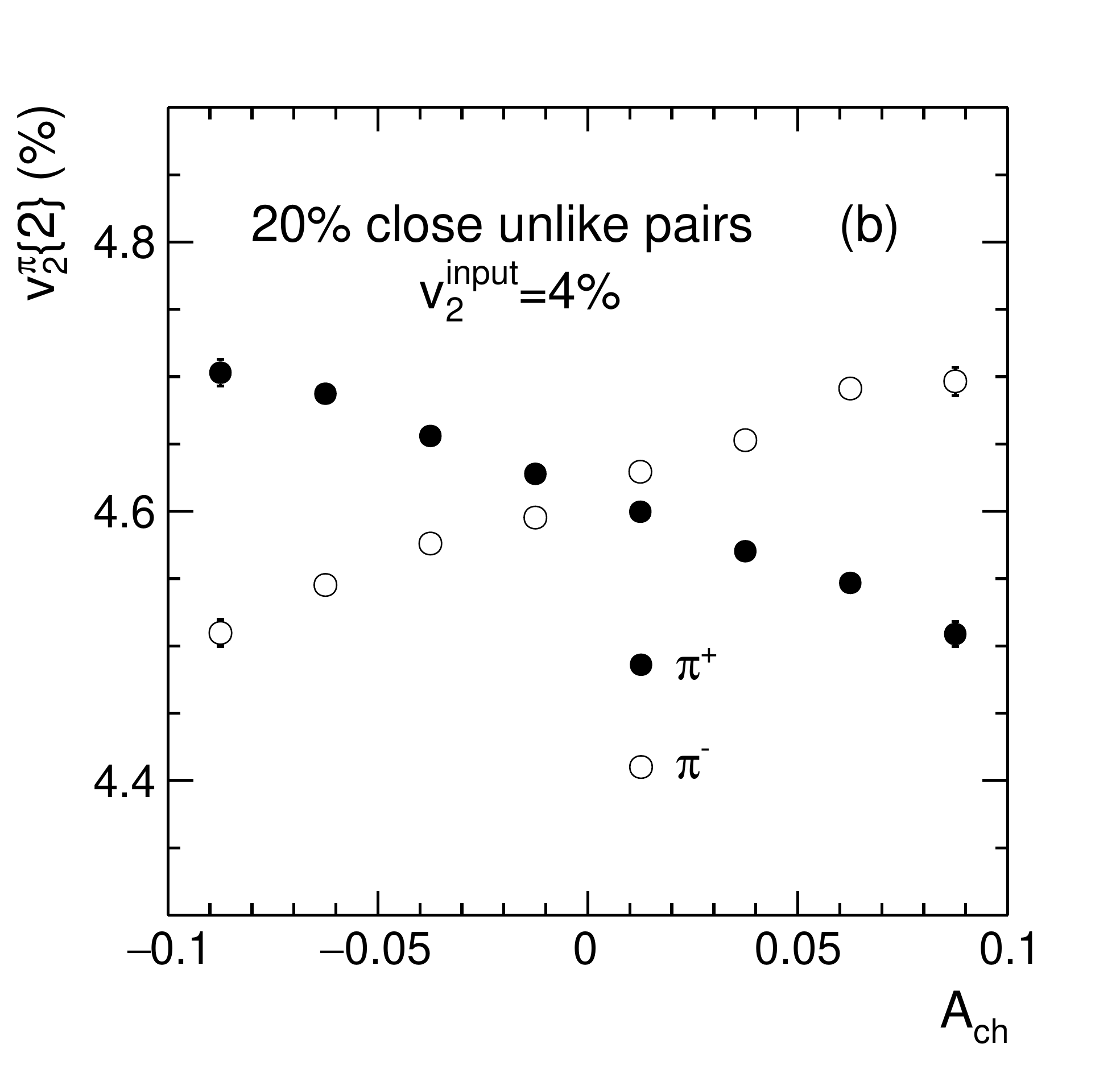} \\
	\includegraphics[scale=0.3399]{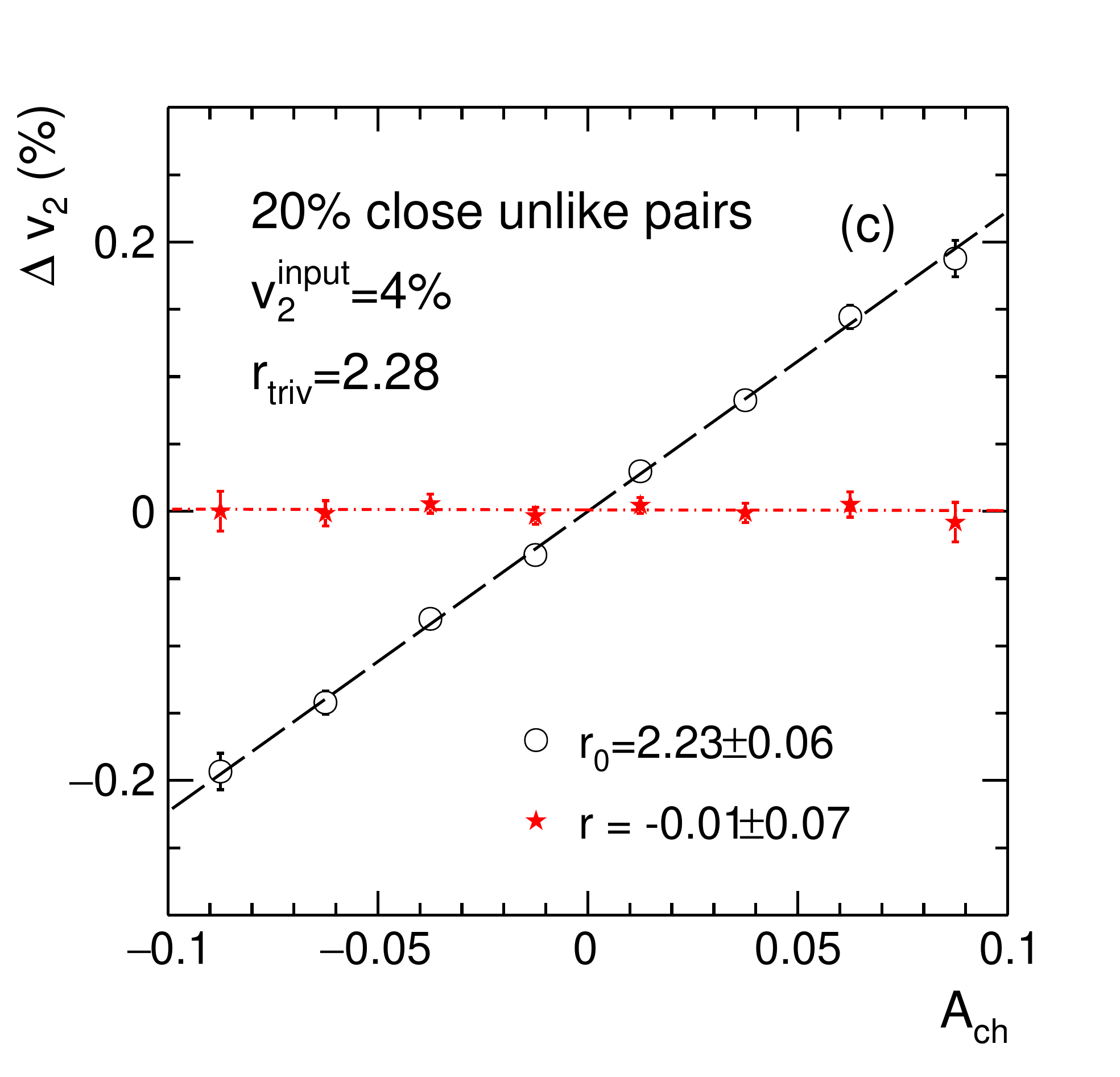} \includegraphics[scale=0.3399]{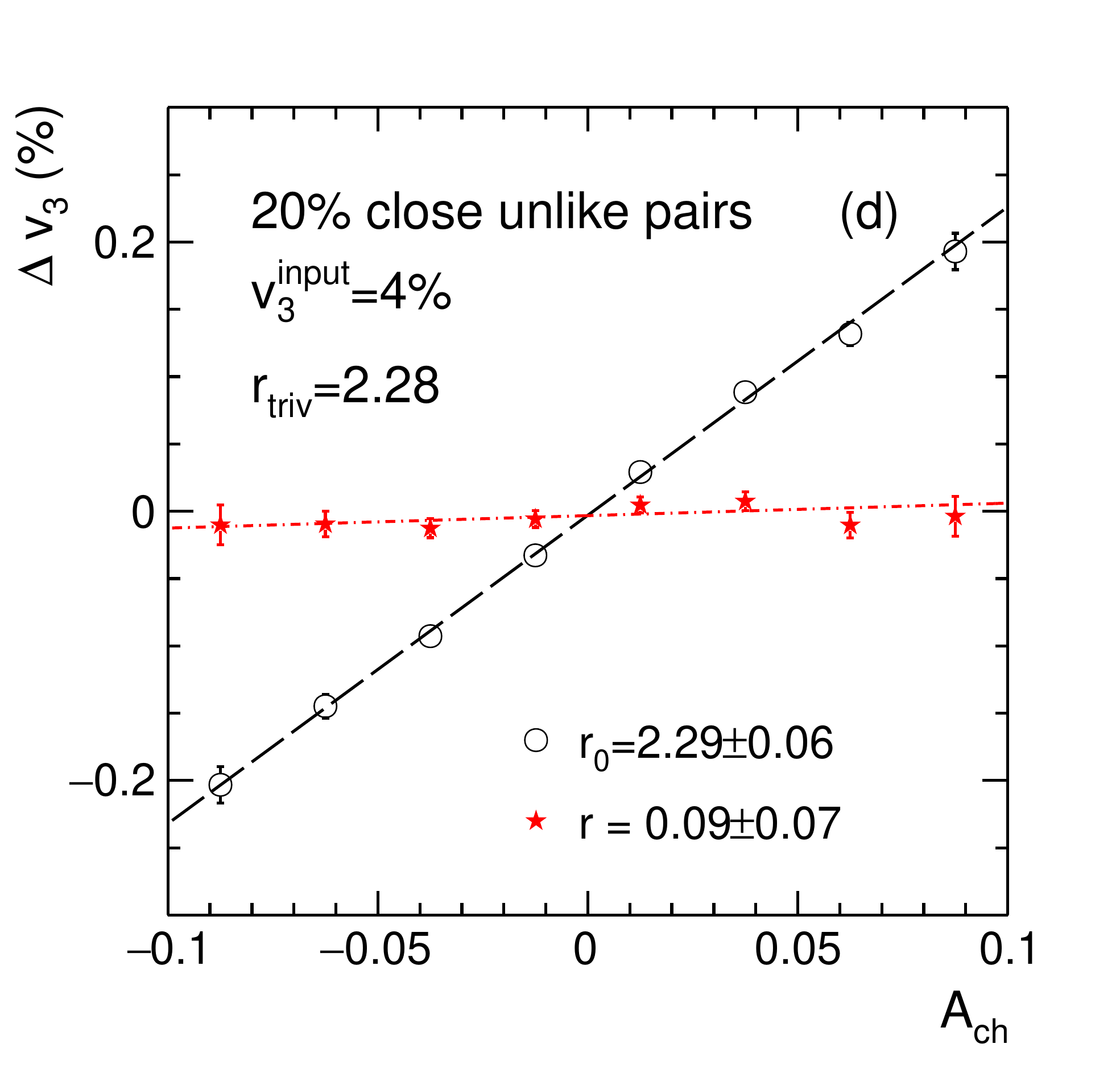}
\par\end{centering}
\caption{(Color online) 
	A toy model demonstration of the trivial linear-$\ach$ term due to the net effect of non-flow 
	difference between like-sign and unlike-sign pairs and using all charged particles as REF.
	(a) The two-particle cumulants $d_{2}\{2\}$ between $\pi^{+}$ and all charged
		hadrons ($h$), positive charges ($h^{+}$) and negative charges ($h^{-}$), respectively. 
	(b) The elliptic flow of $\pi^{\pm}$ calculated from the Q-cumulant method with all charge hadrons as REF. 
	(c) The slope parameters calculated with all charged hadron references ($r_{0}$) and single sign charge references. The $r_{\rm triv}$ is calculated by Eq.~\ref{eq:rT} (i.e, from the trivial term in Eq.~\ref{eq:trivial}).
	(d) Similar to (c) but for $v_{3}$. The non-flow is simulated by unlike-sign close pairs.
	The trivial linear $\ach$ term gives a positive slope to $\dvnach$.
\label{fig:toy1}}
\end{figure*}

Figure~\ref{fig:toy1}(a) shows the two-particle cumulant of $\pi^{+}$ from like-sign ($h^{+}$ as REF), unlike-sign ($h^{-}$ as REF) and all charged hadrons (all charge as REF) correlations.  The like-sign $d_{2}\{2;\pi^{+}h^{+}\}$ is simply $v_{2}^{2}$. The unlike-sign $d_{2}\{2;\pi^{+}h^{-}\}$ is larger because of the input non-flow of the $\pi^{+}\pi^{-}$ pairs.
The elliptic flow of $\pi^{\pm}$ obtained from the $d_{2}$ with all charge REF (black points in Fig.~\ref{fig:toy1}(a) for $\pi^{+}$ POI)  are shown as function of $\ach$ in Fig.~\ref{fig:toy1}(b). The magnitudes are larger than
the input value of $4\%$ because of the input non-flow correlations. These non-flow backgrounds contribute trivial linear-$\ach$ terms to $v_{2}^{\pi^{+}}(\ach)$ and $v_{2}^{\pi^{-}}(\ach)$, with opposite-sign slopes as shown in Fig.~\ref{fig:toy1}(b).
With the input non-flow, we obtain a positive $r_{\rm triv}=2.28\%$ by Eq.~\ref{eq:rT}. This is reflected in the fitted CMW-sensitive slope parameter $r_{0}$, shown in Fig.~\ref{fig:toy1}(c). 

Non-flow differences are present between like-sign and unlike-sign pairs in experiment, and not much can be done to eliminate this non-flow difference.
In order to eliminate the trivial linear $\ach$ term, one is left with only one option, i.e., to  use hadrons of a single charge sign instead of all charged hadrons as REF.
One can use positive and negative particles as REF separately to obtain $v_{n}^{\pi}\{2;h^{+}\}$ and $v_{n}^{\pi}\{2;h^{-}\}$. 
One can then take the average
\begin{equation}
	\bar{v}_{n}^{\pi} \equiv \frac{v_{n}^{\pi}\{2;h^{+}\} + v_{n}^{\pi}\{2;h^{-}\}}{2}.
	\label{eq:avgvn}
\end{equation}
The $\dvvach$ obtained using this technique is shown in Fig.~\ref{fig:toy1}(c) by the red stars.
Indeed, the slope is zero as expected because there is no other physics in our toy model that would introduce a non-zero slope.

The Eq.~\ref{eq:trivial} algebra holds for all orders of azimuthal harmonics.
The trivial $\ach$ term also exists in higher order flow harmonics, e.g, the $\ach$-dependent $\pi$ triangle flow $v_{3}$. 
To illustrate this, we use the same MC model to
generate the azimuth angle using input $v_{3}=4\%$. The results are shown in Fig.~\ref{fig:toy1}(d), and are consistent with those for $v_{2}$.
The underlying reason here is identical for $v_{2}$ and $v_{3}$.

\begin{figure*}[tbp]

\begin{centering}
\includegraphics[scale=0.3399]{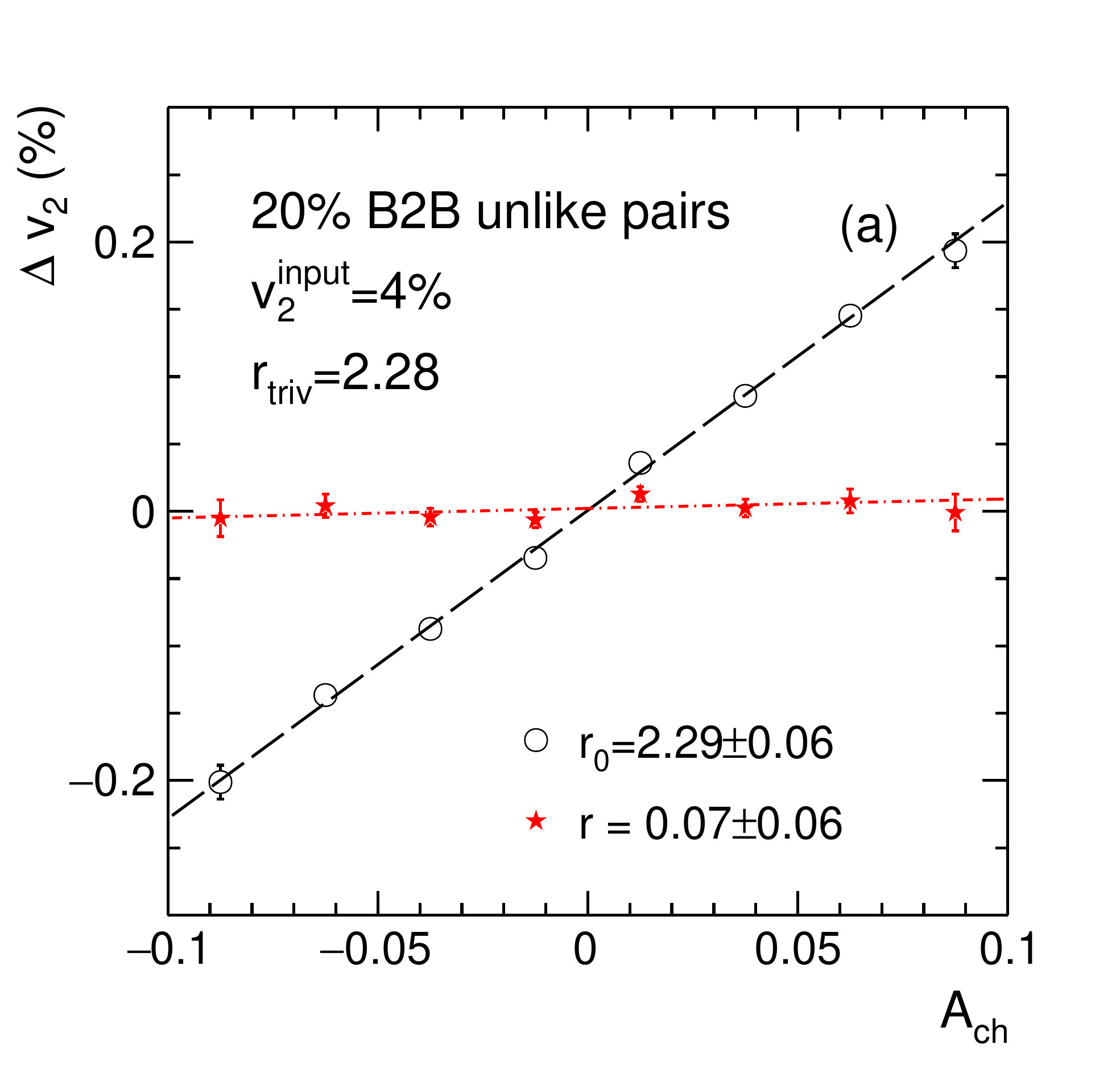}\includegraphics[scale=0.3399]{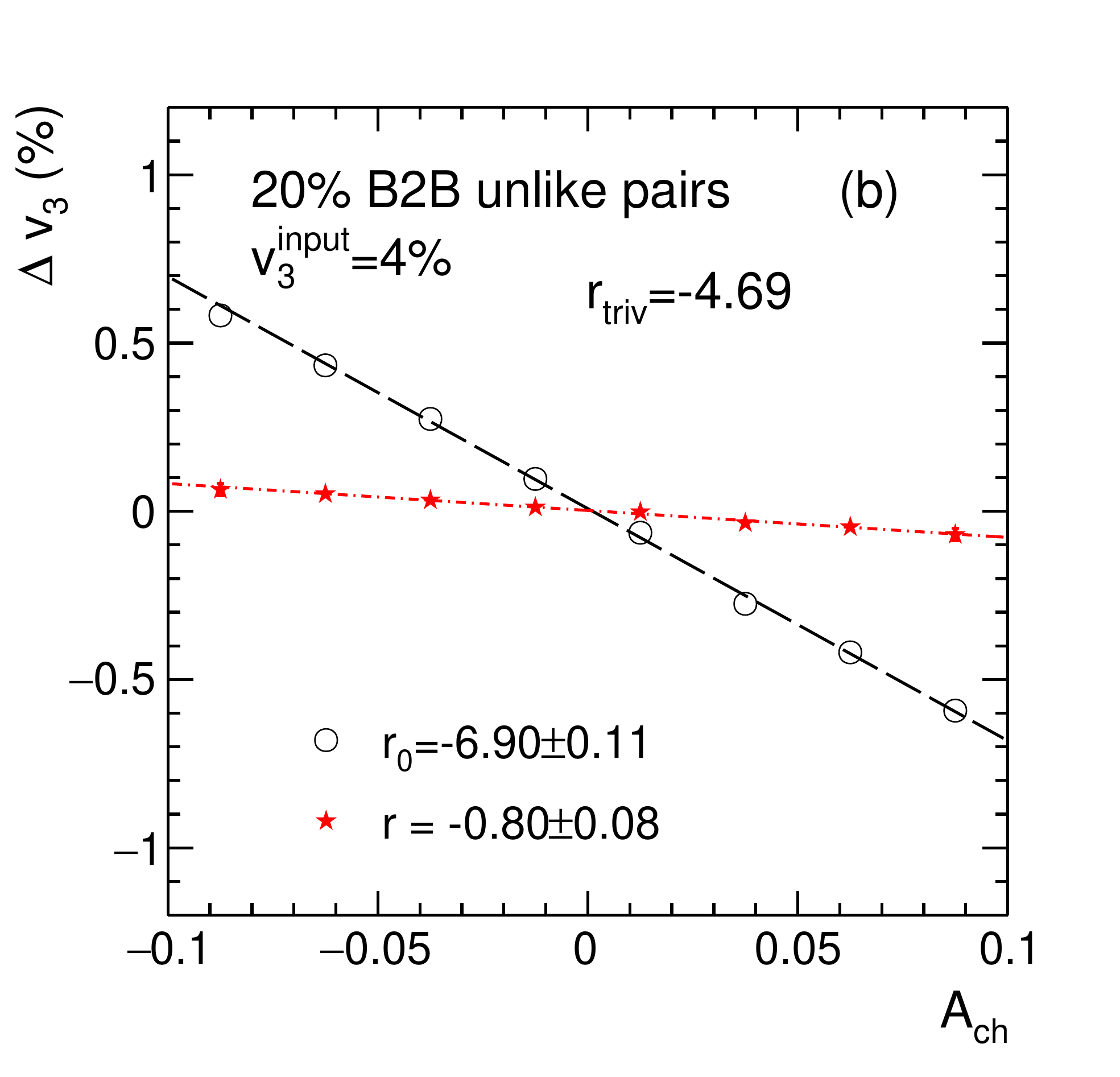} 
\par\end{centering}
	\caption{(Color online) Similar to Fig.~\ref{fig:toy1} but the non-flow is simulated by back-to-back (B2B) unlike-sign pairs. (a) The effect to $\dvvach$ is as same as in Fig.~\ref{fig:toy1}. 
	(b) The effect to $\dvvvach$ is opposite, giving a negative trivial slope.
	The trivial term eliminated slope is still negative.
\label{fig:toy2}}
\end{figure*}

The STAR preliminary results indicate a negative slope for $\dvvvach$ in central and peripheral collisions~\cite{Shou:2018zvw}. The slope in Fig.~\ref{fig:toy1}(d) due to close 
pairs is positive. A negative trivial slope can easily arise from back-to-back pairs because their effect on $v_{3}$ is negative while 
that of close pairs is positive. 
We illustrate this using the same toy model but force the pairs to emit back-to-back. The results are shown in Fig.~\ref{fig:toy2}.
There is no difference between back-to-back pairs and close pairs for $v_{2}$ because they are symmetric with respect to the second harmonic (see Fig.~\ref{fig:toy2}(a) vis-a-vis Fig.~\ref{fig:toy1}(c)). 
While for $v_{3}$, the back-to-back pairs contribute a negative trivial slope as shown in Fig.~\ref{fig:toy2}(b), in contrast to that from close pairs in Fig.~\ref{fig:toy1}(d). 

The trivial term has been removed in $r$, obtained by flow measurements via Eq.~\ref{eq:avgvn}. 
It is interesting to note, however, that the $r$ slope for $\dvvvach$ is non-zero for back-to-back non-flow pairs, as shown in Fig.~\ref{fig:toy2}(b) by the red stars. This is in contrast to the $r$ slopes for $\dvvach$ and that of $\dvvvach$ using close pairs.
The non-zero slope is at first glance hard to understand.
The reason is due to the competition between two sources of pions, the paired pions and unpaired pions, 
which have effectively different final $v_{3}$ values, although both are generated by the same input $v_3$.
This is not the case for $v_{2}$ and $v_{3}$ with the close-pair non-flow.
We will discuss this point in the next section.

To summarize this section, a trivial linear $\ach$ term arises if the following two conditions are both met:
(1) there exists non-flow difference between like-sign and unlike-sign pairs (which is generally true); and 
(2) both charged-sign hadrons are used as REF to calculate flow harmonics, which automatically introduces a $\ach$ dependence. 
Based on our toy model study, we found that for close pair unlike-sign non-flow, the trivial term introduces a positive 
slope to $\dvnach$ for both $n=2,3$.
For back-to-back pair unlike-sign non-flow, the trivial slope is positive for $\dvvach$ and negative for $\dvvvach$.

\section{Linear $\ach$ term due to multiple pion sources}
\label{sec:multiple}

The magnitude of the slope parameter extracted from the $\dvnach$ depends 
on the $\ach$ distribution: a narrower $\ach$ distribution will give a larger slope parameter. In heavy ion collisions,
particle sources with different physics mechanisms  can have different event-by-event fluctuations and thus different$\ach$ distributions. 
For example, the trivial statistical distribution of net-charges is a Skellam distribution, 
while the net-charge distribution with exact charge conservation is a Delta distribution. 

Using a two-component model, i.e. primordial pions (denoted by subscript `P' in the following) and pions from resonance decays (denoted by `D'),
we show that the multiple pion sources introduce a natural $\ach$ dependence in $\dvnach$,
if these pion sources are different in both the $\ach$ distributions and the pion $v_{2}$ magnitudes. 
From  the two-component model, we have 
\begin{eqnarray}
	v_{n\pm} &=&\frac{N_{P\pm}v_{n,P\pm}+N_{D\pm}v_{n,P\pm}}{N_{D\pm}+N_{P\pm}}, \\
	\ach  &=& \frac{A_{P} + \epsilon A_{D}}{1+\epsilon}, \\
	\epsilon &\equiv&\frac{N_{D+}+N_{D-}}{N_{P+}+N_{P-}}, 
\end{eqnarray}
where $N_{P\pm}$ ($N_{D\pm}$) and $v_{n,P\pm}$ ($v_{n,D\pm}$) are the
multiplicity and elliptic flow of primordial (decay) $\pi^{\pm}$,
$A_{P}=(N_{P+}-N_{P-})/(N_{P+}+N_{P-})$, and
$A_{D}=(N_{D+}-N_{D-})/(N_{D+}+N_{D-})$. 

We first assume $v_{n,P/D} = v_{n,P/D+} = v_{n,P/D-}$ and are independent of charge asymmetry.
Then we have 
\begin{align}
	\Delta v_{n} &=\frac{2\epsilon}{(1+\epsilon)^{2}(1-\ach^{2})}(A_{D}-A_{P})(v_{n,P}-v_{n,D}) \nonumber \\
	&  \simeq \frac{2\epsilon}{(1+\epsilon)^{2}}(A_{D}-A_{P})(v_{n,P}-v_{n,D}). \label{eq:twocomponent}
\end{align}
For the sake of simplicity, we assume the event-by-event distributions of $A_P$ and $A_D$ are both normal distributions, 
i.e. $\mathcal{N}(\mu_{P},\sigma_{P}^{2})$  and $\mathcal{N}(\mu_{D},\sigma_{D}^{2})$. Then the event-by-event distribution of $\ach$ 
is also a normal distribution with $\mu_{A} = (\mu_{P}+\mu_{D}\epsilon)/(1+\epsilon)$ and $\sigma_{A}^{2}=(\sigma_{P}^{2}+\sigma_{D}^{2}\epsilon^{2})/(1+\epsilon)^{2}$. 
For a charge-neutral system, we have $\mu_{A}=\mu_{P}=\mu_{D}=0$. Then from Eq.(\ref{eq:twocomponent}) we arrive at
\begin{equation}
	\Delta v_{n}=\frac{2\epsilon(\epsilon\sigma_{D}^{2}-\sigma_{P}^{2})(v_{n,P}-v_{n,D})}{(1+\epsilon)(\epsilon^{2}\sigma_{D}^{2}+\sigma_{P}^{2})}\ach\equiv r^{2C}\ach.
\end{equation}
The slope $r^{2C}$ from the two-component (2C) model is clearly non-zero if $\sigma_{P}^{2}\neq\epsilon^{2}\sigma_{D}^{2}$ and $v_{n,P}\neq v_{n,D}$.
In more general cases, the linear-$\ach$ term is also present in $v_{n,P\pm}$ and $v_{n,D\pm}$, i.e.,
\begin{equation}
	v_{n,P/D\pm} = v_{n,P/D} \mp \frac{r_{P/D}}{2}A_{P/D}.
\end{equation}
In this case, $v_{n,P/D}\equiv(v_{n,P/D+} + v_{n,P/D-})/2$ and 
\begin{equation}
	\Delta v_{n}\simeq \left.(\frac{\epsilon^{2}\sigma_{D}^{2}r_{D}}{\epsilon^{2}\sigma_{D}^{2} + \sigma_{P}^{2}} + \frac{\sigma_{P}^{2}r_{P}}{\epsilon^{2}\sigma_{D}^{2} + \sigma_{P}^{2}} + r^{2C}\right.)\ach.
	\label{eq:twocomp}
\end{equation}
Therefore, besides the linear-$\ach$ term
from $v_{n,P\pm}$ and $v_{n,D\pm}$ themselves, the competition between two pion sources introduces another
linear-$\ach$ term. The root reason is that the relative fractions of pions from different sources depend on the event-by-event $\ach$ value (because they contribute to $\ach$ differently),
therefore the average $v_{2}$ from multiple sources, which have different $v_{2}$'s, will depend on $\ach$.

We have used two ``flow'' sources in the above derivation. However, this also applies to the competition between flow and non-flow
contributions to the observed $\dvnach$. 
This is the reason for the non-zero slope in Fig.~\ref{fig:toy2}(b) obtained from back-to-back pairs,
because the ``$v_{3}$'' from the single pion flow and the ``$v_{3}$'' from the back-to-back pair non-flow are different,
even though the single pions and the back-to-back pairs are generated with the same $v_{3}$.
Such a problem is not present for $v_{2}$. 
We have tested $v_{2}$ using two different input $v_{2}$ for single and paired pions, 
and also found a non-zero slope parameter.

\section{Non-flow background beyond the trivial contribution}
\label{sec:likesign}
After addressing the trivial term by using REF of single charge sign, we now focus on the trivial slope eliminated results.
In the rest of the paper, we only discuss the physical slope parameter $r$.

\begin{figure}[tbp]

\begin{centering}
\includegraphics[scale=0.3399]{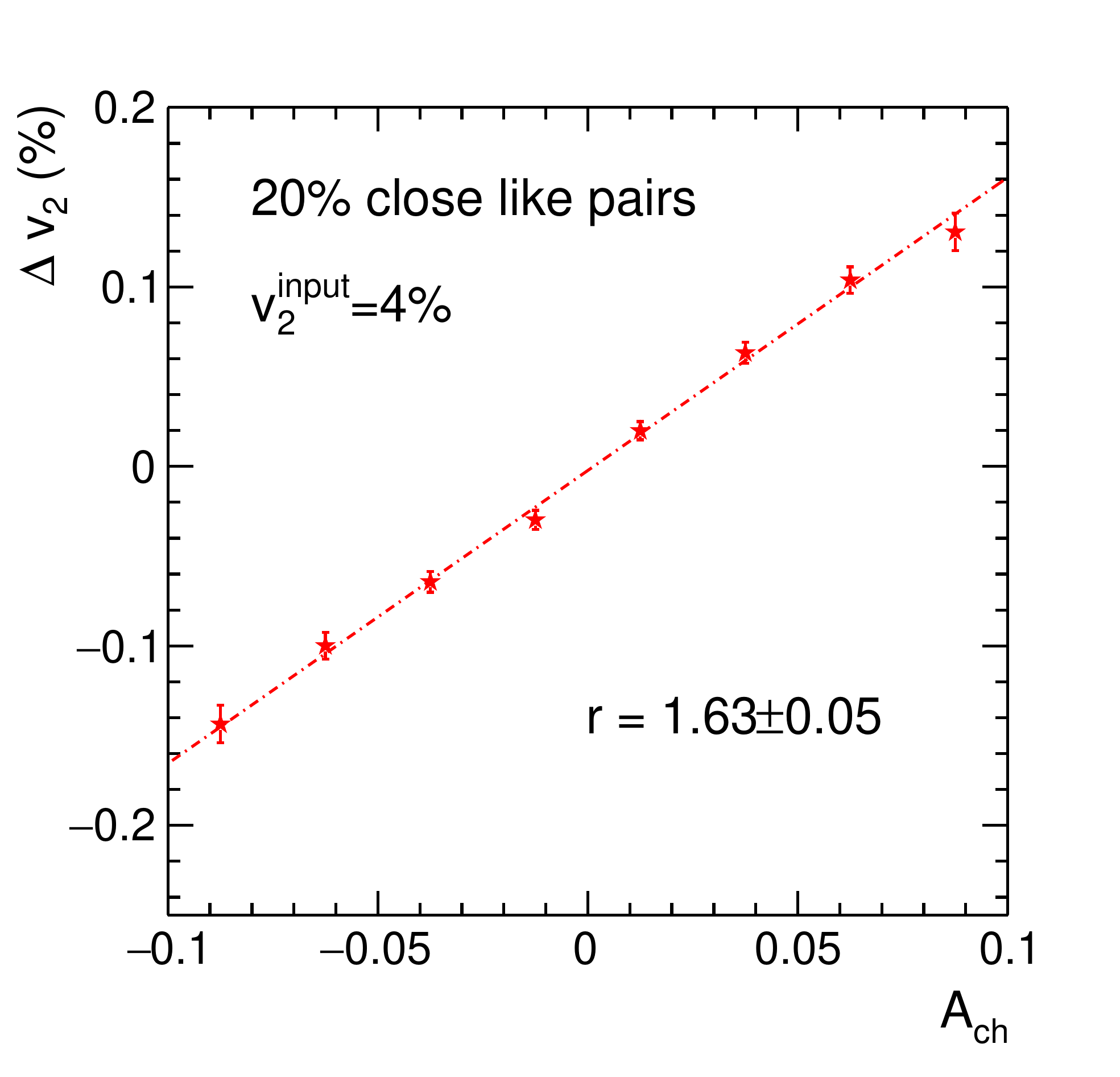}
\par\end{centering}
	\caption{(Color online) 
	Effect of like-sign non-flow correlations on the slope of $\dvvach$.
\label{fig:toy4}}
\end{figure}

As shown in Sec.~\ref{sec:trivial}, the physical slope parameter $r$ of $\dvvach$ is not affected by unlike-sign non-flow correlations. However, as we will show now, 
the like-sign non-flow correlations still introduce a nonzero slope parameter.
We modify our non-flow toy model to generate like-sign pairs instead of unlike-sign pairs. 
We force $20\%$ 
of $\pi^{+}$ (and $\pi^{-}$) to be paired in $\pi^{+}\pi^{+}$  (and $\pi^{-}\pi^{-}$) with the same azimuth.
The results are shown in Fig.~\ref{fig:toy4}.
The  $\dvvach$ has a positive slope $r=1.63\%$. 
This is due to the dilution effect: when more $\pi^{+}$ are counted resulting in a positive $\ach$, the $\pi^{+}\pi^{+}$ non-flow correlation is more diluted while 
the $\pi^{-}\pi^{-}$ non-flow is less diluted. This is different in the unlike-sign correlations discussed before, because in that case, the 
dilution depends on the number of $\pi^{+}\pi^{-}$ pairs. The dilution effect for unlike-sign non-flow is almost identical for $\pi^{+}$ and $\pi^{-}$.

\section{Effect of decay kinematics}
\label{sec:kinematics}
As we have mentioned in the introduction, local charge conservation can also yield a non-zero slope of $\dvnach$~\cite{Bzdak:2013yla}.
Effect of the LCC mechanism on $\dvnach$ comes from the resulting pion correlations and the $p_{T}$-dependent pion flow.
Specifically, the charge asymmetry
is more affected by neutral clusters with lower $p_{T}$, and thus smaller $v_{2}$~\cite{Bzdak:2013yla}. 
A lower $p_{T}$ parent resonance has a larger decay opening angle. The decay daughters have a larger probability to cross acceptance boundaries to cause a 
finite $\ach$, and those decay daughters have smaller $v_{2}$. Moreover, if the parent $v_{2}$ decreases with increasing $|\eta|$, the effect would be even stronger. 

\begin{figure*}

\begin{centering}
\includegraphics[scale=0.3399]{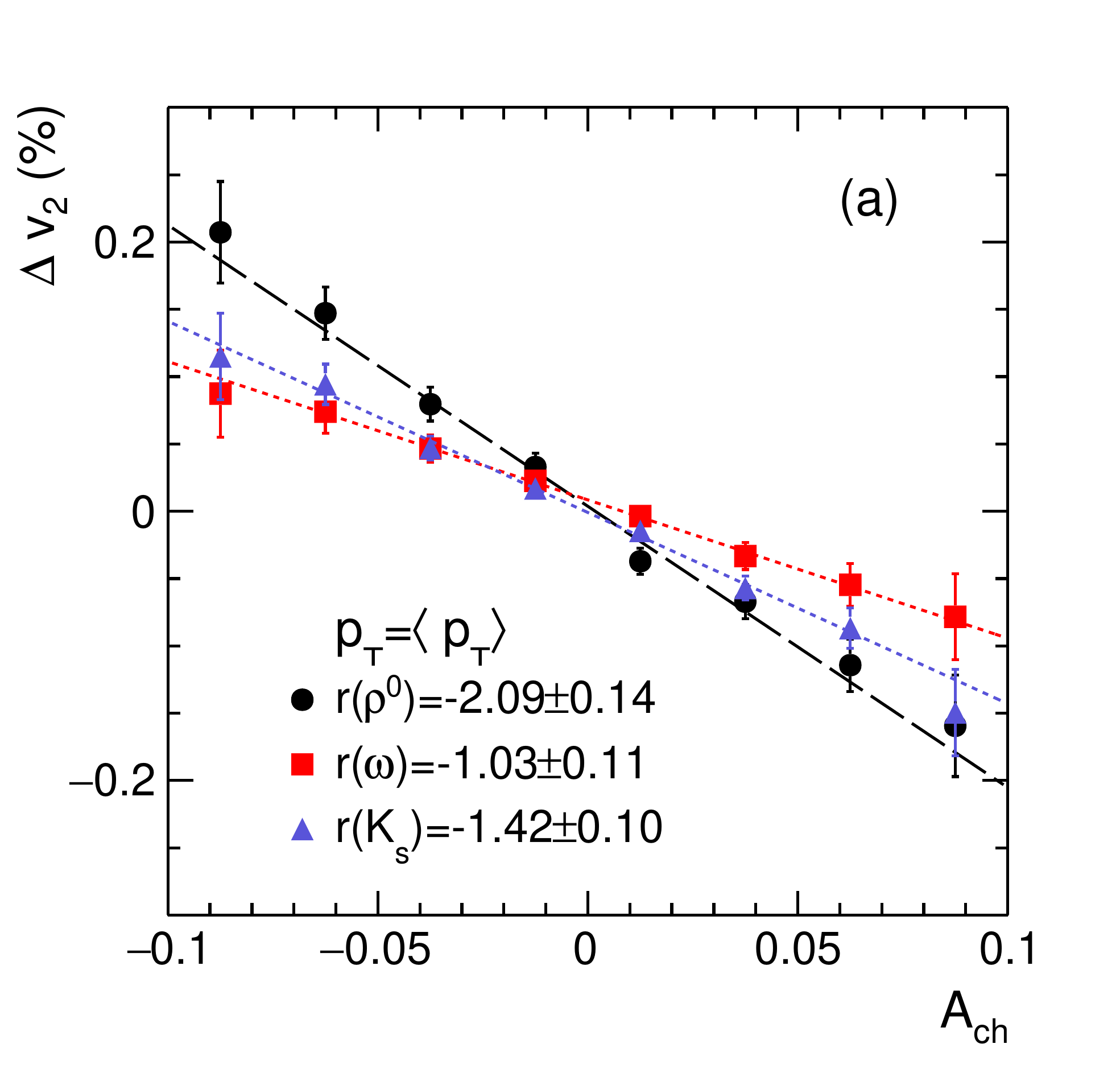}
\includegraphics[scale=0.3399]{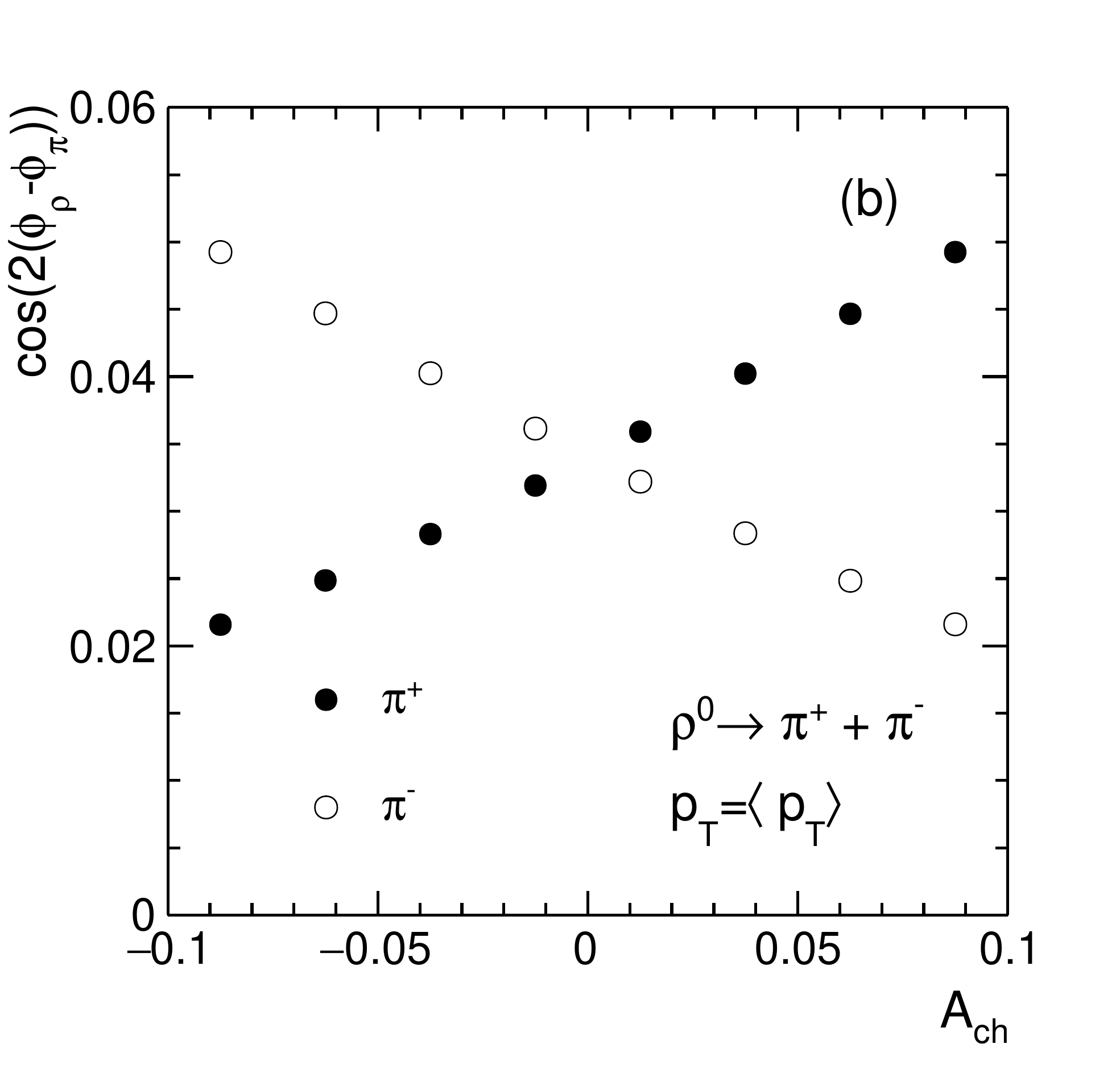} \\
\includegraphics[scale=0.3399]{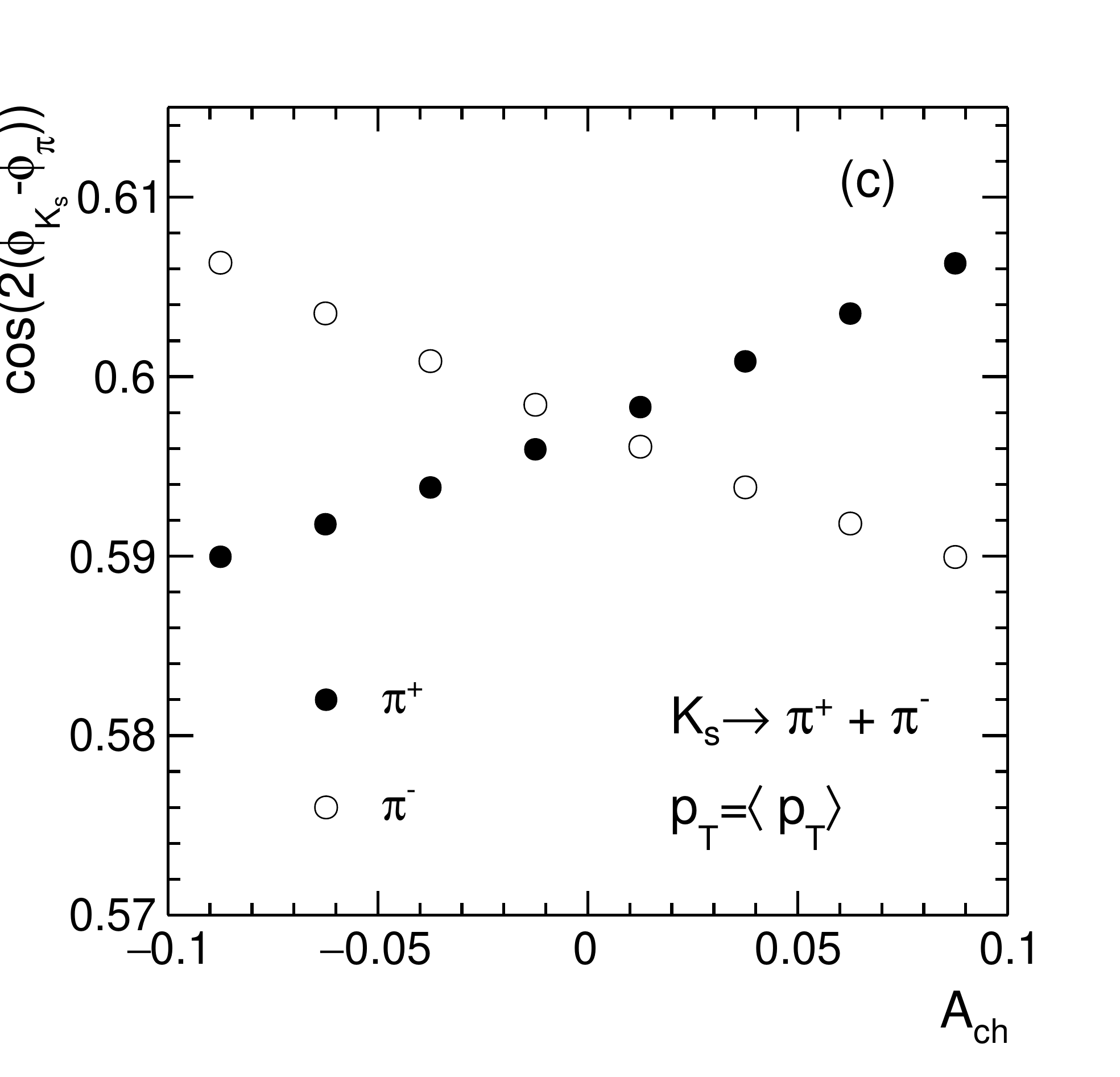}
\includegraphics[scale=0.3399]{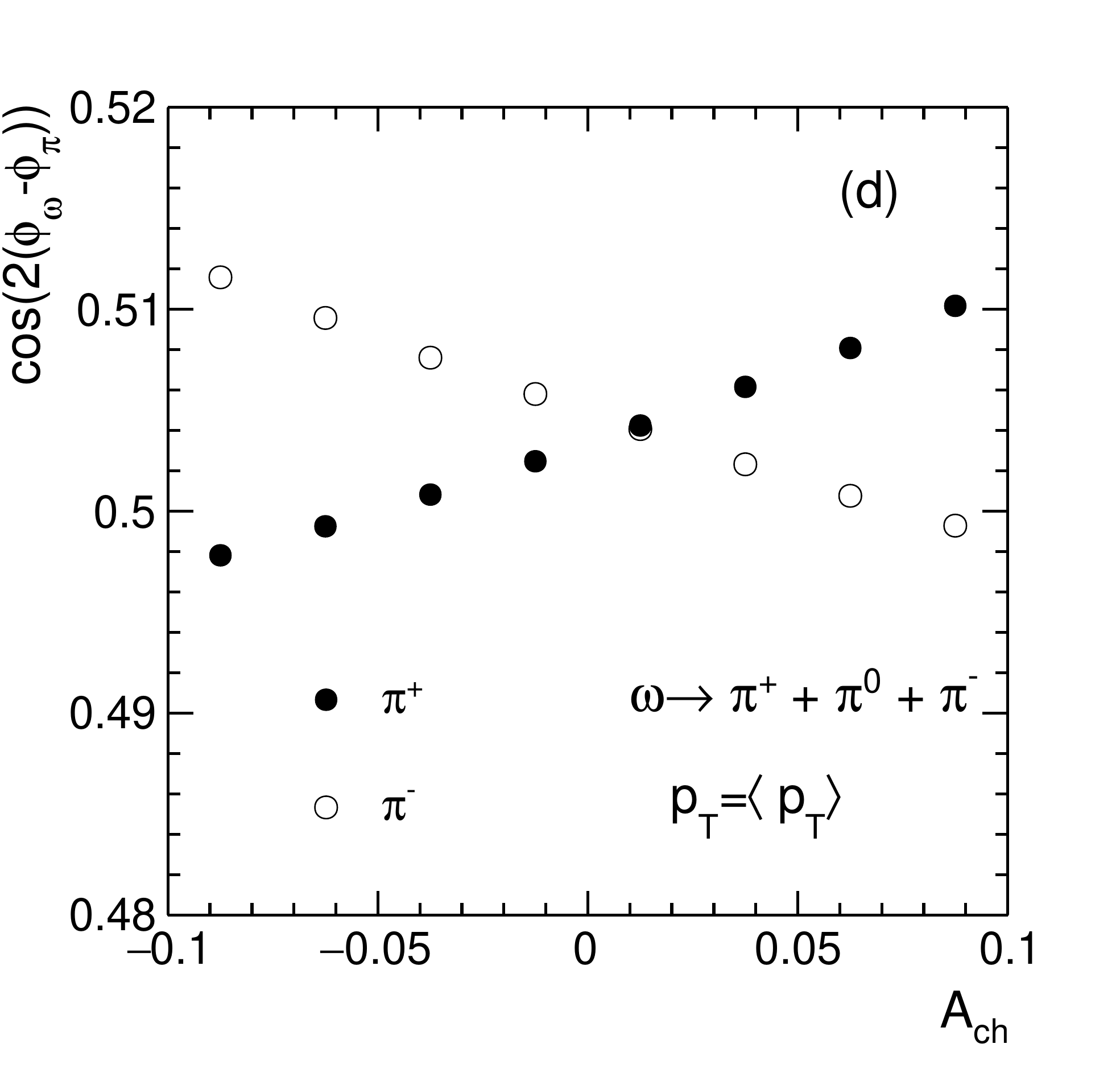} \\
\par\end{centering}
	\caption{(Color online) (a) The $\pi$ elliptic flow differences $\dvvach$ from a single source of $\rho^{0}$, $K_{s}$, or $\omega$ decays with fixed $p_{T}=\langle p_{T}\rangle$ of the parent; 
	The azimuth angle correlation between the parents  (b)$\rho^{0}$, (c)$K_{s}$, (d) $\omega$ and their decay daughter pions. 
	The elliptic flows are calculated from like-sign correlations. The acceptance cut for pions is $0.15<p_{T}<0.5$ GeV/c and $|\eta|<1$. 
	\label{fig:decay}}
\end{figure*}

\begin{figure*}[tbp]

\begin{centering}
\includegraphics[scale=0.3399]{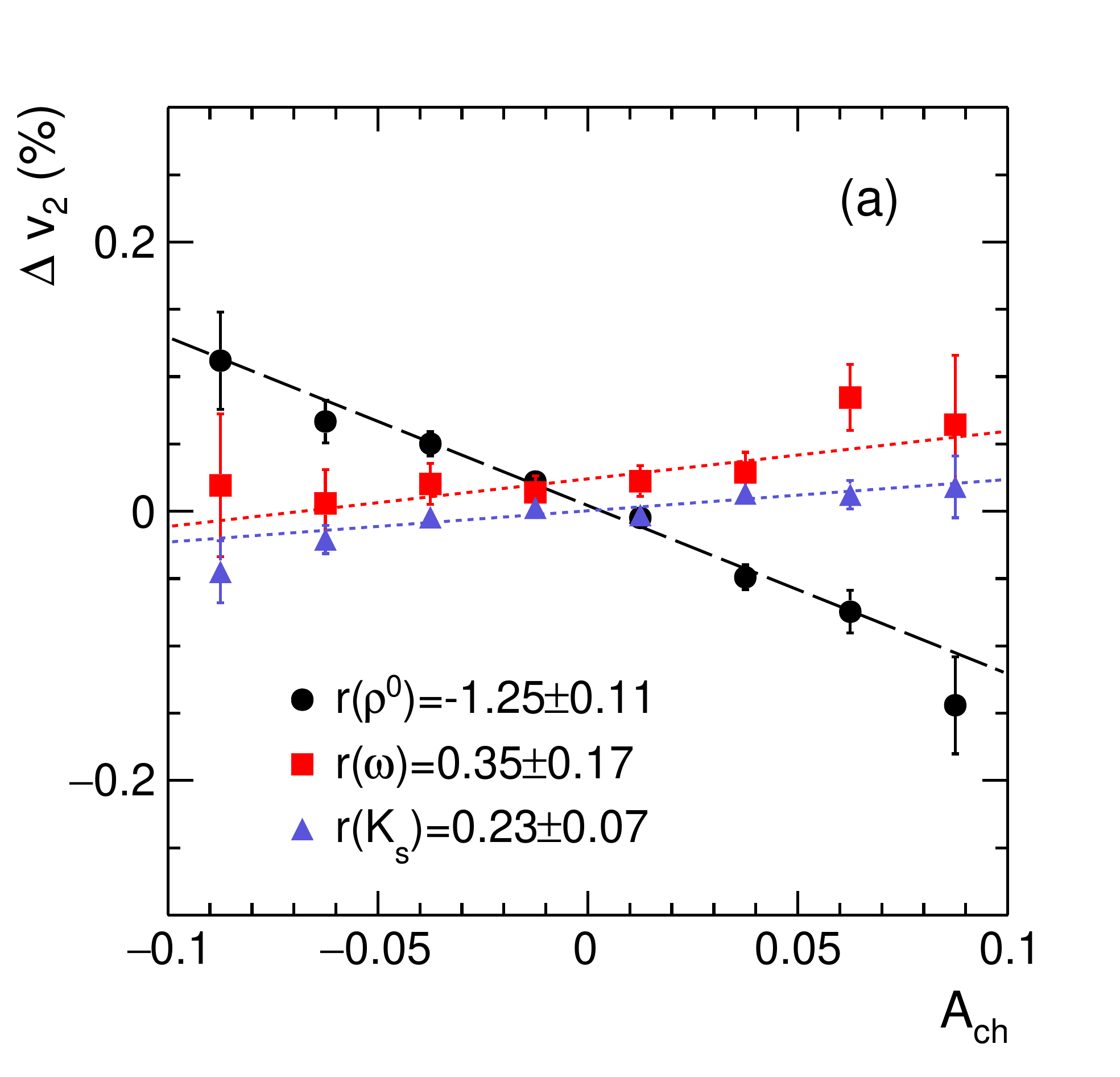}
\includegraphics[scale=0.3399]{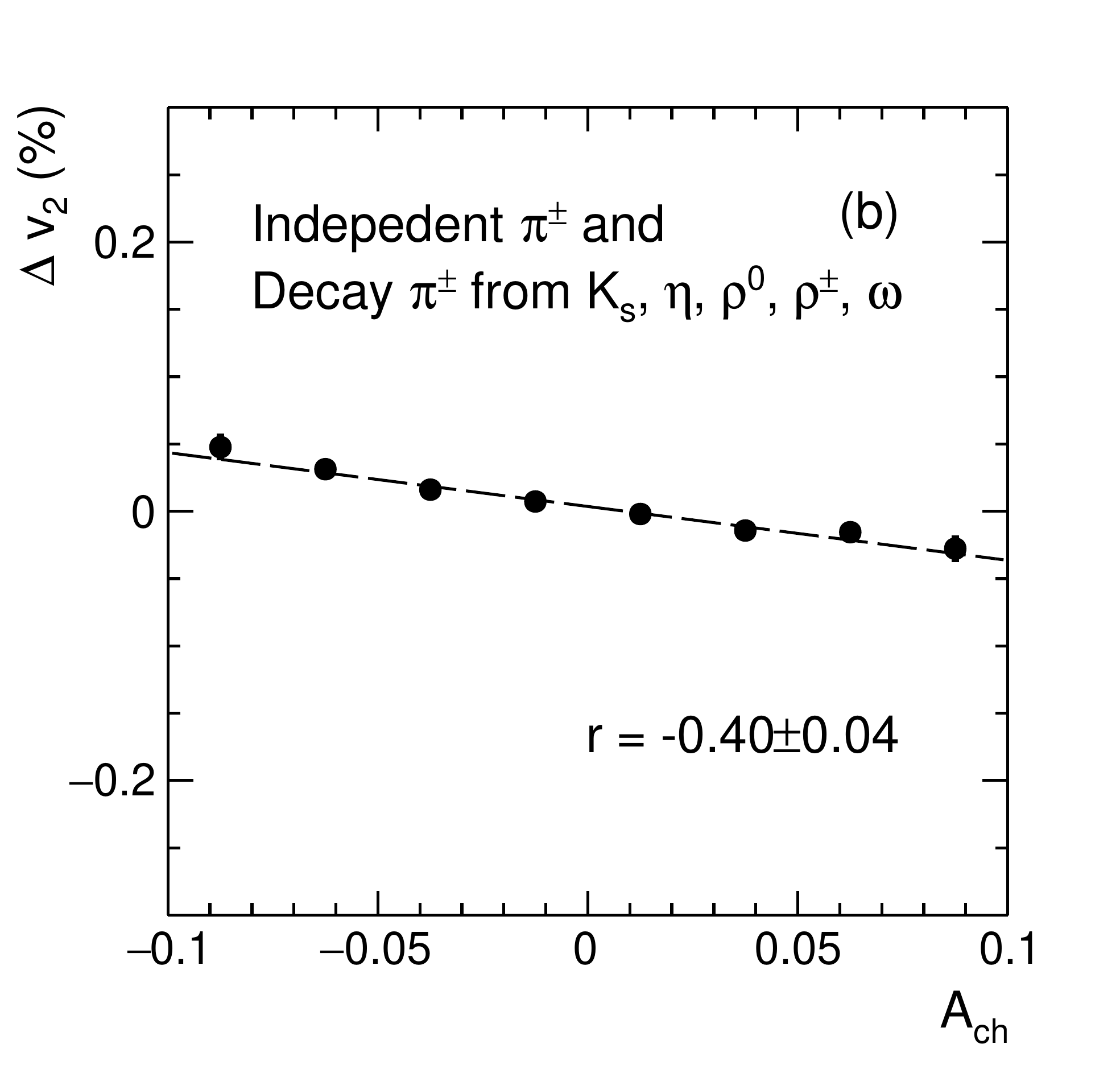} 
\par\end{centering}
	\caption{(Color online) The $\pi$ elliptic flow differences $\dvvach$ from (a) a single source of $\rho^{0}$, $K_{s}$, or $\omega$ decays, and 
	(b) multiple pion sources, primordial pions 
	and decay pions from $K_{s}$,$\rho^{0},\rho^{\pm},\omega,\eta$, with the primordial to decay pion multiplicity ratio within acceptance of $0.94$. 
	The elliptic flows are calculated from like-sign correlations. The acceptance cut for pions is $0.15<p_{T}<0.5$ GeV/c and $|\eta|<1$. 
	\label{fig:multdecay}}
\end{figure*}

\begin{figure*}[tbp]
\begin{centering}
\includegraphics[scale=0.3399]{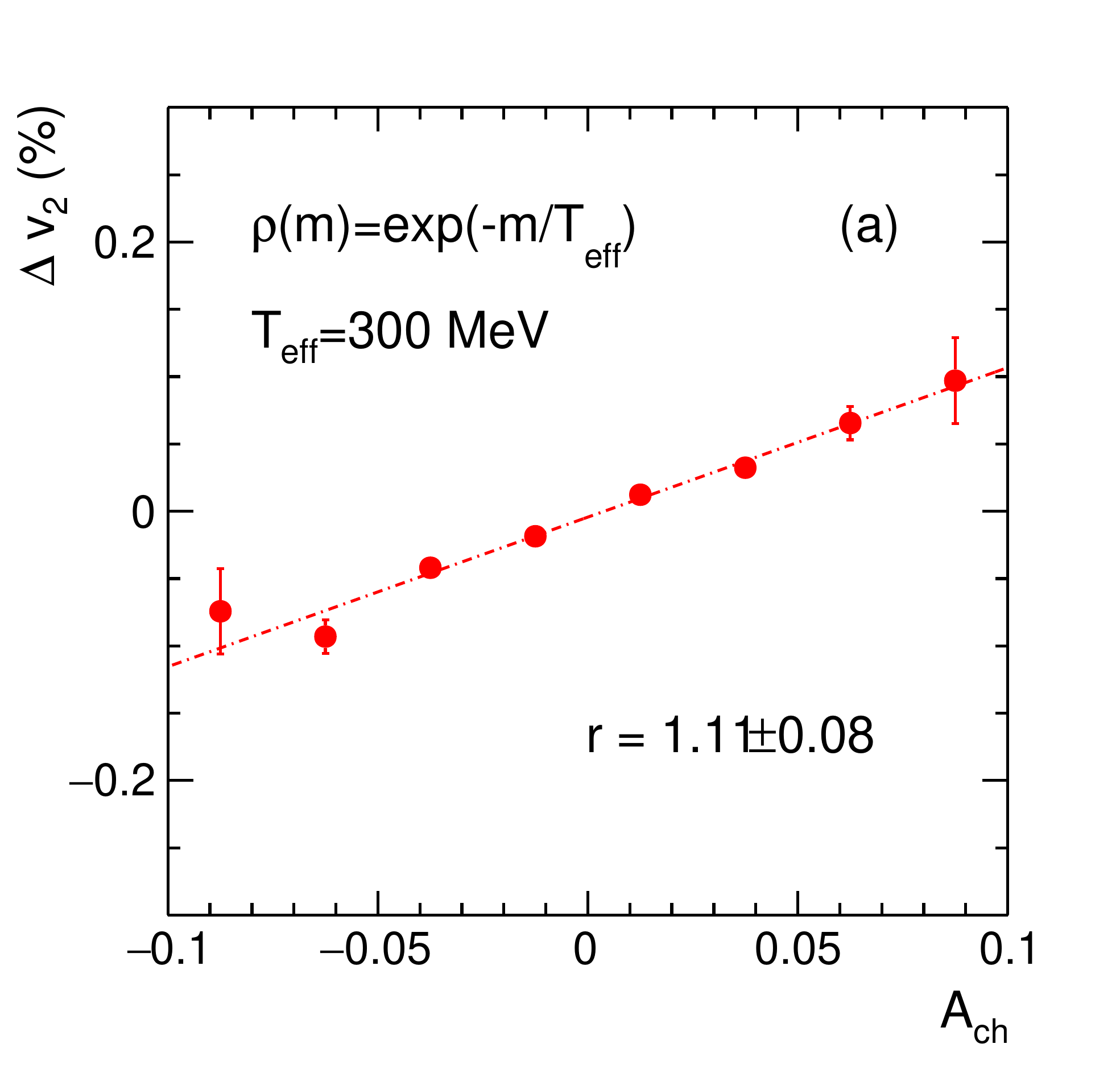}
\includegraphics[scale=0.3399]{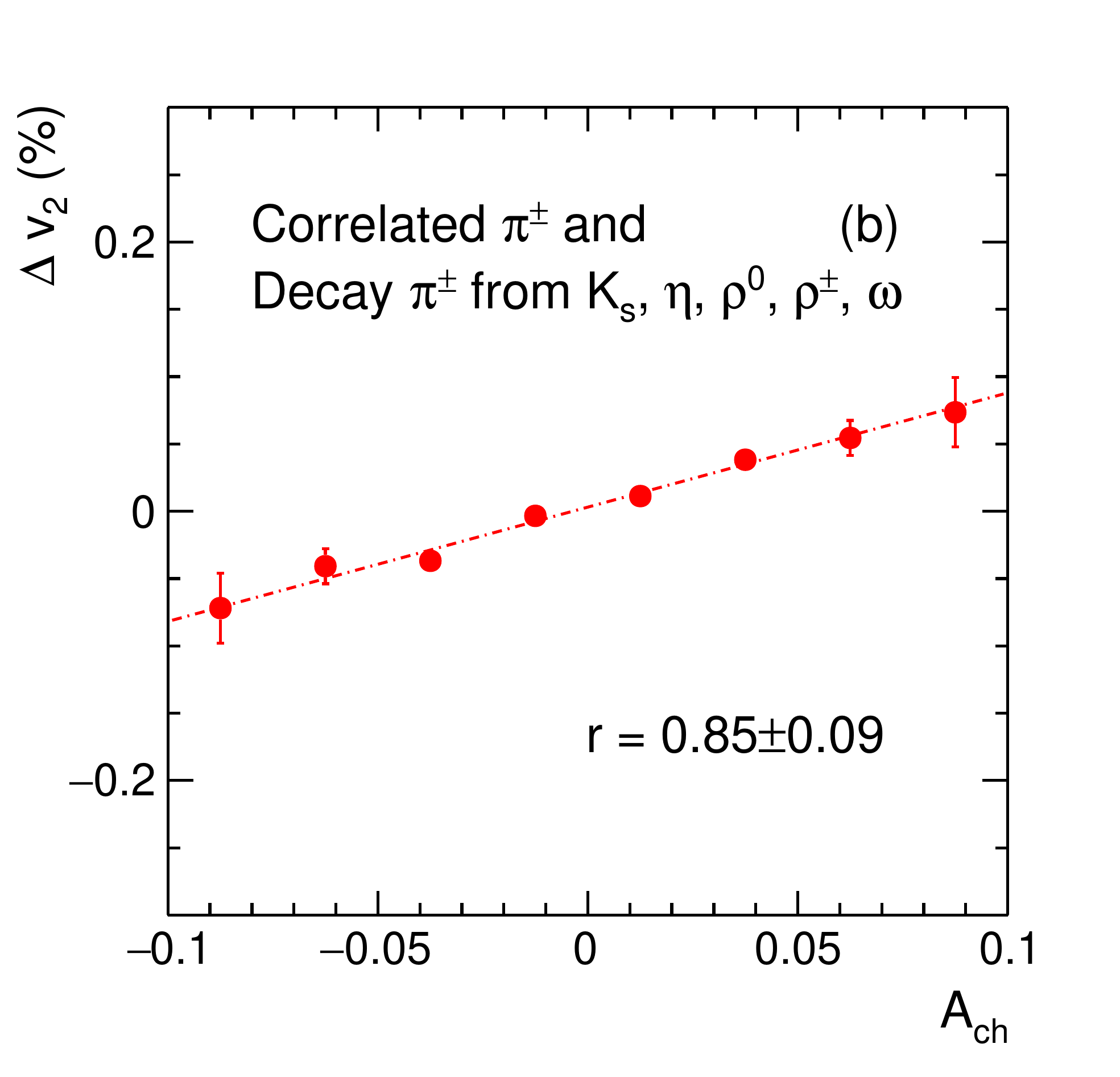}
\par\end{centering}
	\caption{(Color online) 
	(a) $\dvvach$ from decays of neutral clusters with continuum invariant mass spectrum. 
	(b) Similar to Fig.~\ref{fig:multdecay}(b) but use the correlated primordial pions from Panel (a) to replace the independent pions.
	The elliptic flows are calculated from like-sign correlations. The acceptance cut for pions is $0.15<p_{T}<0.5$ GeV/c and $|\eta|<1$. 
	\label{fig:continuum}}
\end{figure*}

In principle, the trivial linear-$\ach$ term discussed in Sec.~\ref{sec:trivial} can be eliminated by using any kind of single-sign reference, but the non-flow
backgrounds could be different. The resonance decays only contribute the unlike-sign non-flow correlations. To simplify our discussion of LCC,
in the rest of this paper, the elliptic flow is calculated from the like-sign correlations to avoid the non-flow contamination.  Namely, the slope parameter $r$ is then extracted from
\begin{equation}
	\dvnach\equiv v_{n}^{\pi^{-}}\{2;h^{-}\} - v_{n}^{\pi^{+}}\{2;h^{+}\}.
\end{equation}

In this section, we use a MC model to study the effect of resonance decays on $\dvvach$. 
We first study the effect of resonance decays with the parent resonances at fixed $p_{T}=\langle p_{T}\rangle$ (see below the related $p_T$ spectra) 
and a pseudorapidity independent resonance flow, so that the aforementioned $p_{T}$-dependent LCC effect is absent.
The averaged multiplicity of total charges in the STAR acceptance ($p_{T}>0.15$ GeV/c, $|\eta|<1.0$) is about $380$ with the $\eta$ spectra given in Eq.~\ref{eq:eta}. 
The $v_{2}$ is kept uniform in $\eta$.
The slope $r$ of $\dvvach$ from $\rho^{0}\rightarrow\pi^{+}+\pi^{-}$, $K_{s}\rightarrow\pi^{+}\pi^{-}$~\footnote{Although $K_s$ is a stable particle, we will use the term ``resonance'' everywhere for convenience.}, and $\omega\rightarrow\pi^{+}+\pi^{0}+\pi^{-}$ 
are shown in Fig.~\ref{fig:decay}(a) individually.
The slopes are all negative. This is because of the following. For resonances within the acceptance,
those decay products closer to the parent (smaller angular difference from the resonance) are more likely to stay within the acceptance.
For resonances outside the acceptance, the decay daughters further away more likely cross into the acceptance.
Because of the nonlinear mapping between the decay angle and the acceptance gauge of $\eta$,
the net effect is that the daughter particles that are accepted within a given $\eta$ window have, on average,
smaller angle from the parent than those outside the acceptance.
This results in a less angular smearing of the daughter particle flow.
This is shown in Fig.~\ref{fig:decay}(b-d) by the azimuthal 
angle correlations between parent ($\rho^{0}$, $K_{s}$, $\omega$) and its decay pions as a function of $\ach$. 
At positive $\ach$, more $\pi^{+}$ are within the acceptance and they have a smaller average opening angle from the parent.
The excess pions due to $\ach$ selection will contribute a larger elliptic 
flow, thus the negative slope for $\dvvach$.

\section{Effect of local charge conservation}
\label{sec:lcc}
We now include the $p_{T}$ distributions to the above analysis to study the LCC effects in our MC simulation. 
The $p_{T}$ spectra, and $v_{2}(p_{T})$ of $\rho^{0}$, $K_{s}$, and $\omega$ we used correspond to the measured data in $30\mbox{-}40\%$ centrality of Au+Au collisions~\cite{Zhao:2017nfq,Adamczyk:2015lme}. 
The average multiplicity of total charges in the STAR acceptance ($p_{T}>0.15$ GeV/c, $|\eta|<1.0$) is about $380$ with the $\eta$ spectra given in Eq.~\ref{eq:eta}. 
The $v_{2}$ is kept uniform in $\eta$.
The results are shown in Fig.~\ref{fig:multdecay}(a). 
For the $\rho^{0}$ meson, the LCC contribution is small, and the slope parameter $r$ is still negative. 
This is because the decay opening angle is large for the relevant $p_{T}$ range
where the $\rho^{0}$ yields are appreciable. The feeding into and out of the acceptance of the daughter pions are rather insensitive to the $p_{T}$ of the $\rho^{0}$.
For decays with a smaller mass, like $K_{s}$, the decay opening angle is relatively small, so the positive contribution from LCC is appreciable.
This makes the final slope positive. 
Three-body decays usually have small decay opening angles, so the LCC effect is also appreciable. 
Indeed, positive slope parameter is observed from $\omega$ decays.
In general, smaller mass resonance decay gives a large positive slope due to the LCC effect.

We apply the above simulations to multiple sources of pions, namely, decays of $K_{s}$, $\rho^{0}$, $\rho^{\pm}$, $\omega$, and $\eta$, together with primordial pions.  
The parameters are taken from Ref.~\cite{Zhao:2017nfq,Adamczyk:2015lme} and the primordial to decay pion multiplicity ratio within acceptance is taken to be $0.94$.
The mean multiplicity of total charged pions within STAR acceptance is $380$ with the $\eta$ distribution given in Eq.~\ref{eq:eta}.
Poisson multiplicity fluctuations are applied for all the particle species. 
The result is shown in Fig.~\ref{fig:multdecay}(b). 
The slope parameter is about $-0.4\%$, 
where contribution from the competition among multiple pion sources discussed in Sec.~\ref{sec:multiple} is also included.

The negative slope parameter from the multiple pion source simulation is different from the one observed 
by Ref.~\cite{Bzdak:2013yla}, where a positive slope was obtained  from a simplified MC simulation and a 3+1-dimension hydrodynamic calculation.
The reason is that, besides the pions from resonance decays, the primordial $\pi^{+}\pi^{-}$ are also considered as correlated pairs by the LCC mechanism in Ref.~\cite{Bzdak:2013yla}. 
The $(p_{T},\eta)$ correlations of these primordial $\pi^{+}\pi^{-}$ pairs 
are constructed while the azimuthal
angles are sampled from the input $v_{n}$ modulation.
In our case, the azimuthal angles of decay pions are given by the parent azimuth and decay kinematics, see Fig. ~\ref{fig:decay}(b-d).
Furthermore, in the MC and hydrodynamic calculation in Ref.~\cite{Bzdak:2013yla} the LCC is implemented as neutral cluster decays with 
the mass spectrum, $p_{T}$ spectrum, $v_{2}(p_{T})$, and decay kinematics  
different from the known resonances discussed here.
One marked difference is the continuum-like  invariant mass spectrum of those neutral clusters.
To illustrate this effect, we use an exponential mass spectrum,
\begin{equation}
	\rho(m) = \exp(-m/T_{\rm eff}),
	\label{eq:continuum}
\end{equation}
with $T_{\rm eff}= 300$ MeV to mimic the invariant mass spectrum obtained by sampling pairs from the single pion $p_{T}$ spectrum keeping the $\eta$ the same.
The correlations are introduced by the two-body decay mechanism.
For the sake of comparison, the $p_{T}$ spectrum, $v_{2}(p_{T})$ of neutral clusters, and the mean multiplicity 
of the decayed $\pi$'s in the STAR acceptance are as same as those used in the above $\rho^{0}$ study. 
The result is shown in Fig.~\ref{fig:continuum}(a).
A positive slope parameter $r=1.11\%$ is obtained from this mass spectra. 
This is in-line with the above observation that smaller resonance mass gives
a more positive slope because the average mass of the spectrum  of Eq.~\ref{eq:continuum} is small.
We note that the effect of the convex $v_{2}(\eta)$ distribution discussed in Ref.~\cite{Bzdak:2013yla} is not included in our study, which will further increase the slope parameter.

We apply the LCC correlated pions from neutral clusters of the continuum mass
to replace the independent primordial pions in the above multiple source simulations. The results are shown in Fig.~\ref{fig:continuum}(b) with the slope $r\sim0.85\%$. 
The result differs from the one shown in Fig.~\ref{fig:multdecay}(b). The reason are twofold: First, the slope parameter of $\dvvach$ generated from correlated pions is not zero, as we 
have shown in Fig.~\ref{fig:continuum}(a), while the one from independent primordial pions is obviously zero. Second, the multiplicity distribution of net-charge from
correlated pions is much narrower than the Skellam distribution from independent primordial pions. This will give a different contribution to 
the slope of $\dvvach$ from the competition among multiple pion sources.

It is worthwhile to note that the propose of our study is not to give quantitative descriptions of the measured data 
but to illustrate the potential non-CMW mechanisms contributing to
the CMW-sensitive slope parameter. More elaborate studies are needed to pin down the exact LCC contribution to the slope observable.
Considering that the CMW-sensitive observable is also sensitive to kinematics and multiplicity fluctuations, more extensive background studies are called for.

\section{Summary}

The charge asymmetry ($\ach$) dependent pion elliptic flow difference $\dvvach$ was proposed as a sensitive observable to the chiral magnetic wave (CMW) search.
In this paper, we first demonstrated that the flow measurements can automatically introduces a linear-$\ach$ dependence if 
(1) there exists non-flow difference between like-sign and unlike-sign pairs and 
(2) both charged-sign hadrons are used as reference in two-particle cumulants to calculate the flow harmonics.
Based on our toy model study, we found that for close pair unlike-sign non-flow, the trivial term introduces a positive trivial 
slope to $\dvnach$ for both $n=2,3$.
For back-to-back pair unlike-sign non-flow, the slope is positive for $\dvvach$ and negative for $\dvvvach$.   

After eliminating the trivial term, 
we further study other non-CMW mechanisms that can produce non-zero slopes in $\dvvach$. We found:
 \begin{itemize}
	 \item Non-flow between like-sign pairs gives a positive slope to $\Delta v_{2}(A_{\rm ch})$ 
because of the larger dilution effect to $\pi^{+}$ ($\pi^{-}$) at positive (negative) $A_{\rm ch}$. 
\item Competition among multiple $\pi$ sources can introduce a linear-$\ach$ term.
	This effect is sensitive to the differences in multiplicity fluctuations and anisotropic flows of those sources, and arises from the $\ach$-dependent relative contributions of pions from those sources. 
Such $\ach$ dependence does not need any $\ach$ dependence of the elliptic flow of each individual $\pi$ source.
	\item The slope parameter is sensitive to the kinematics of neutral clusters decaying into pions.
	Light resonances give positive slopes while heavy resonances give negative slopes. 
		 Local charge conservation (LCC) from continuum cluster mass spectra can give positive slope parameters.

 \end{itemize}

Our studies indicate that many physics mechanisms can give rise to a $A_{\rm ch}$ dependent $\Delta v_{2}(A_{\rm ch})$. 
One could produce a positive slope of a few present, as experimentally observed, from those non-CMW mechanisms.
The interpretation of $\Delta v_{2}(A_{\rm ch})$ in terms of CMW is therefore delicate. 
In order to identify the CMW, those (and potentially more) non-CMW mechanisms have 
to be assessed in detail and to a precision that would leave no doubt to an experimental signal.
This would require accurate modeling of particle/resonance production and dynamics of heavy ion collisions.

\section*{Acknowledgments}
This work is supported in part by the National Natural Science Foundation of 
China (Grant Nos. 11905059, 11847315) and the U.S. Department of Energy (Grant No. de-sc0012910).  
HX acknowledges financial support from the China Scholarship Council.

\bibliography{ref}
\end{document}